# On the Time Series Length for an Accurate Fractal Analysis in Network Systems

G. Millán, *Member, IEEE*

*Abstract*— It is well-known that fractal signals appear in many fields of science. LAN and WWW traces, wireless traffic, VBR resources, etc. are among the ones with this behavior in computer networks traffic flows. An important question in these applications is how long a measured trace should be to obtain reliable estimates of de Hurst index (*H*). This paper addresses this question by first providing a thorough study of estimator for short series based on the behavior of bias, standard deviation ($\sigma$), Root-Mean-Square Error (RMSE), and convergence when using Gaussian *H*-Self-Similar with Stationary Increments signals (*H-sssi* signals). Results show that Whittle-type estimators behave the best when estimating *H* for short signals. Based on the results, empirically derived the minimum trace length for the estimators is proposed. Finally for testing the results, the application of estimators to real traces is accomplished. Immediate applications from this can be found in the real-time estimation of *H* which is useful in agent-based control of Quality of Service (QoS) parameters in the high-speed computer network traffic flows.

## I. Introducción

LOS procesos estadísticamente autosimilares representan el comportamiento estocástico invariante a cambios de escalas [1]-[4]. Estos procesos se aplican ampliamente como modelos en una amplia gama de campos de la ciencia [5]-[10]. En el área de las redes de computadoras estos procesos se usan para modelar el tráfico LAN agregado, el tráfico de vídeo VBR, el tráfico inalámbrico y el tráfico debido al WWW, entre otros. En todos estos estudios el tráfico se mide y luego analiza con la finalidad de determinar si se ajusta o no a un comportamiento autosimilar. Las trazas de tráfico utilizadas en estos estudios se componen de cientos de miles de puntos y la obtención de tales longitudes de las muestras suele implicar largos tiempos de captura. Ocurre que para estudios de tráfico fuera de línea [11] dichas longitudes y tiempos pueden ser aceptables, mientras que para aplicaciones de administración de red en tiempo real de métricas de QoS basadas en la estimación precisa de *H* [12], dichas longitudes y sus tiempos de captura son terminantemente inaceptables. Este artículo primero estudia el comportamiento de los estimadores a series temporales de corta longitud y luego aborda el problema de obtener la longitud mínima requerida para las estimaciones exactas en tiempo real de *H*. Por lo tanto el requisito es obtener una alta precisión con una longitud mínima en contraposición con lo planteado en [13] donde se plantea que no es posible determinar una longitud mínima para las series temporales autosimilares sin que se pierdan sus propiedades intrínsecas. La exactitud debe ser comparable con la de series largas, en las cuales la precisión se basa para este caso en métricas tales como $\sigma$, sesgo (BIAS) y RMSE. El análisis de convergencia constituye también una herramienta útil para lo anterior. Luego, en este artículo se aborda el siguiente problema: dadas la(s) especificación(es) de sesgo, $\sigma$ o MSE, ¿cuál debe ser la longitud mínima, $N_{min}$ de la serie temporal que las satisfaces?; es decir, supóngase que el proceso estocástico $\psi$ posee un índice de Hurst $H_\psi$, se pide encontrar la longitud mínima $N_{min}$ de la serie temporal tal que para cada realización de longitud $N_{min}$ los índices de Hurst estimados, $H_{\psi e}$, son similares a $H_\psi$. Para lograrlo el artículo se organiza de la siguiente manera. La sección II introduce de forma breve los conceptos básicos de procesos estocásticos y los métodos para estimar el grado de fractalidad, *H*. En la sección III se describe la metodología para encontrar $N_{min}$, mientras que la sección IV estudia con detalle el comportamiento de los estimadores de *H* en series de corta longitud y el problema de hallar la magnitud mínima de dichos estimadores. En la sección V se aplican los resultados a trazas de tráfico reales y, finalmente en la sección VI se entregan las principales conclusiones del trabajo realizado.

## II. Fractalidad, Autosimilitud y Estimación de *H*

Un proceso fractal o autosimilar es aquel cuya distribución de probabilidades es invariante a las dilataciones temporales y la compresión de amplitud [14].

Sea $Z = \{Z_t\}_{t \in I}$, donde $I = \mathbb{R}$ o $\mathbb{R}_+$, un proceso estocástico real valuado. Se dice que *Z* es fractal si y solo si existe $H \in \mathbb{R}$ tal que para todo $a \in \mathbb{R}_+$ se cumple que $\{Z_{at}\}_{t \in I} =_d \{a^H Z_t\}_{t \in I}$, donde $=_d$ significa igualdad en el sentido de sus distribuciones de probabilidad. Generalmente el interés está en los procesos fractales con incrementos estacionarios para los cuales todo lo anterior se cumple con $H > 0$ [15]. La definición anterior se llama estricta. Una versión relajada se obtiene mediante la definición de segundo orden, la cual requiere invariancia en las propiedades estadísticas de segundo orden. Formalmente, Sea $Z_t$ un proceso estocástico de tiempo continuo. Se dice que $Z_t$ es autosimilar de segundo orden si y solo si $E(Z_t) = a^{-H} E(Z_{at})$ y $R_{zz}(t,s) = a^{-2H} R_{zz}(at, as)$. Ya que las redes de computadoras requieren modelos de tiempo continuo se necesitan versiones discretas de todo lo anterior. Sea $X = \{X_t, t \in \mathbb{Z}\}$ un proceso de tiempo discreto, posiblemente obtenido por muestreo de una señal aleatoria de tiempo continuo. *X* es estrictamente autosimilar si y solo si $X =_d m^{1-H} \Gamma_m(\{X\})$ con $H \in (0,1)$ para todo *m* natural donde $\Gamma(\cdot)$ proceso de agregación de bloques que recibe por entrada una serie temporal de longitud *N* y

G. Millán, Departamento de Ingeniería Eléctrica, Universidad de Santiago de Chile, ginno.millan@usach.cl.

entrega una serie temporal de longitud $N/m$ [16]. Una versión relajada de la definición anterior es la de autosimilitud de segundo orden en el sentido exacto. Formalmente, $X$ es un proceso exactamente autosimilar de segundo orden si [17]

$$\rho(k) = \frac{1}{2}[(k+1)^{2H} - 2k^{2H} + (k-1)^{2H}]. \quad (1)$$

Un proceso que tiene una función de correlación de la forma anterior también satisface [18]

$$\text{Var}(X) = m^{2-2H}\text{Var}(\Gamma_m(\{X\})), \text{ y} \quad (2)$$

$$\text{Cov}(\Gamma_m(X_t), \Gamma_m(X_{t+k})) = m^{2-2H}\text{Cov}(X_t, X_{t+k}). \quad (3)$$

En el ámbito de las redes de computadoras se utiliza una versión mucho más relajada de (1): un proceso $X$ se dice que es asintóticamente autosimilar de segundo orden si la función de correlación de $\Gamma_m(\{X\})$ cuando $m \to \infty$ es igual a la de un proceso exactamente autosimilar de segundo orden en tiempo discreto, es decir, es igual a (1). Si en la definición exacta o asintótica de autosimilitud $k \to \infty$, entonces $\rho(k) = ck^{2(1-H)}$, lo cual implica dependencia de largo alcance (LRD). Significa que un proceso autosimilar exacto o asintótico presenta LRD siempre que $H \in (0,1)$ y $k \to \infty$. Cuanto mayor el valor de $H$, más suave será el proceso y más lenta será la descomposición a cero de las autocorrelaciones. Se han propuesto métodos de diversa índole para estimar $H$; estos pueden clasificarse como métodos pertenecientes al dominio del tiempo, al dominio de la frecuencia y métodos tiempo-escala. Entre los métodos de dominio del tiempo se encuentra el estadístico R/S [19], [20], el gráfico de tiempo de varianza, la varianza de los residuos, el momento absoluto, el método de Higuchi, la variación de la ventana a escala [21], MAVAR, Whittle, etc. [22], [23]. GPH, periodograma y otros métodos de periodogramas modificados se encuentran en la clase de dominio de la frecuencia que, a su vez, aprovecha el comportamiento de ley de potencia de los procesos autosimilares en la vecindad del origen. Los métodos tiempo-escala incluyen estimadores basados en wavelets como el estimador Abry-Veitch [24]-[28]. Las herramientas software para el análisis de similitud de también son importantes ya que recopilan una serie de estimadores y metodologías para mejorar el análisis de la autosimilitud.

### III. Metodología

En primer lugar, se proporciona un estudio detallado del comportamiento de los estimadores a series cortas y luego se proponen longitudes mínimas para dichos estimadores. El comportamiento de los estimadores se estudia aplicando un estimador dado a $N$ series temporales y luego se obtienen $\sigma$, BIAS y RMSE. La comparación del comportamiento de los estimadores sujetos a estos estadísticos y para diferentes longitudes de series, ayuda a encontrar la longitud mínima. También un análisis de convergencia muestra la evolución de los estimadores en el tiempo y será útil en este documento. A continuación, se describen estos pasos con más detalle. Para aplicar los estimadores, se deben obtener $N$ series de tiempo. Las señales sintéticas con $H$ conocido se obtienen mediante la simulación de una serie Gaussiana H-sssi (fGn) utilizando el método de Davies y Harte [9]. Las longitudes consideradas para 100 trazas fueron $N = \{2^i, i = 6,7,...,16\}$ con parámetros de Hurst $H \in \{0.5, 0.6, 0.7, 0.8, 0.8, 0.9, 0.99\}$, es decir se consideró un total de 6600 señales fractales exactas. Para cada conjunto de estimaciones de un parámetro $H$ en particular, se calculan las siguientes estadísticas BIAS $= H_0 - \bar{X}$, donde $H_0$ es el valor nominal de $H$; la desviación estándar $\sigma$ y RMSE tal que RMSE $= N^{-1}\sum_{i=1}^{N}(x_i - H_0)^2$. Basado en los resultados de BIAS, $\sigma$ y RMSE se propone una longitud mínima $N_{min}$. $N_{min}$ se obtiene a partir de estimaciones precisas (BIAS $\sim 0.03$ y $\sigma \sim 0.015$). Además, también se propone la clasificación de las estimaciones basadas en los valores de BIAS y $\sigma$. Se obtienen estimaciones de alta precisión cuando BIAS $\sim 0.03$ y $\sigma \sim 0.015$, estimaciones aceptables cuando BIAS $\in (0.03, 0.05)$ y $\sigma \le 0.02$ y estimaciones sesgadas cuando BIAS $> 0.1$. Una vez que se obtiene la longitud mínima para las series de tipo fGn, la aplicación de estos resultados se realiza con trazas largas tanto sintéticas como reales. Para dicha serie $Z$ de longitud $M$, $M > N_{min}$, se estudia lo siguiente: sea $t_0, t_1,..., t_k$ una secuencia de puntos en el eje $X$, donde $t_{i+1} > t_i$ y $(t_{i+1} - t_i) < N_{min}$, a cada bloque de $Z$ de longitud $N_{min}$, $\{Z_j\}_{j=t_i}^{t_i+N_{min}-1}$ se aplica un método de estimación del parámetro de Hurst $\hat{H}_{t_i}^{N_{min}}(\cdot)$. Lo anterior se repite hasta que $t_k + N_{min} > M$ para cualquier $k$. Toda vez que $N_{min}$ se encuentre correctamente elegido, un gráfico $t_i$ versus $\hat{H}_{t_i}^{N_{min}}(\cdot)$ debería resultar en una señal con poca variación (esta variación debería ser igual a $\sigma$). La longitud $N_{min}$ propuesta se relaciona con el análisis de convergencia de una serie, motivo por el cual también se estudia como sigue. La convergencia de cualquier estimador se obtiene dividiendo la serie $Z$ en bloque de tamaño $m \ll M$ para obtener $Z = \{\Psi_1^m, \Psi_2^m,..., \Psi_i^m\}$, dond $\Psi_i^m = \{Z_{(i-1)m}, Z_{(i-m)m+1},..., Z_{im}\}$. Luego se aplica $\hat{H}_{t_i}^{N_{min}}(\cdot)$ para $\bigcup_{j=1}^{N/m}\{\Psi_i^m\}$ para obtener $\hat{H}_1^m, \hat{H}_1^{2m},..., \hat{H}_1^{jm}$. Luego se grafica $\hat{H}_i^{jm}$ versus $jm$ para $j = 1, 2, ..., N/m$ con el cual se verifica el comportamiento del estimador $\hat{H}_{t_i}^{N_{min}}(\cdot)$.

### IV. Simulaciones y Resultados

En la Fig. 1 se presentan dos diagramas en perspectiva de las estimaciones del parámetro de Hurst para trazas con $H = 0.81$ y para longitud variable al aplicar técnicas basadas en wavelet y Whittle. La gráfica izquierda corresponde a la técnica de wavelet mientras que la gráfica derecha a Whittle. Un ejemplo de estas trazas obtenida con SELFIS [15] se muestra en la Fig. 2. Obsérvese en el gráfico de la izquierda de la Fig. 1 que las técnicas basadas en wavelet experimentan sesgo y variabilidad altos al estimar el parámetro $H$ para series temporales breves. La longitud de las trazas es de $N < 2^{12}$.

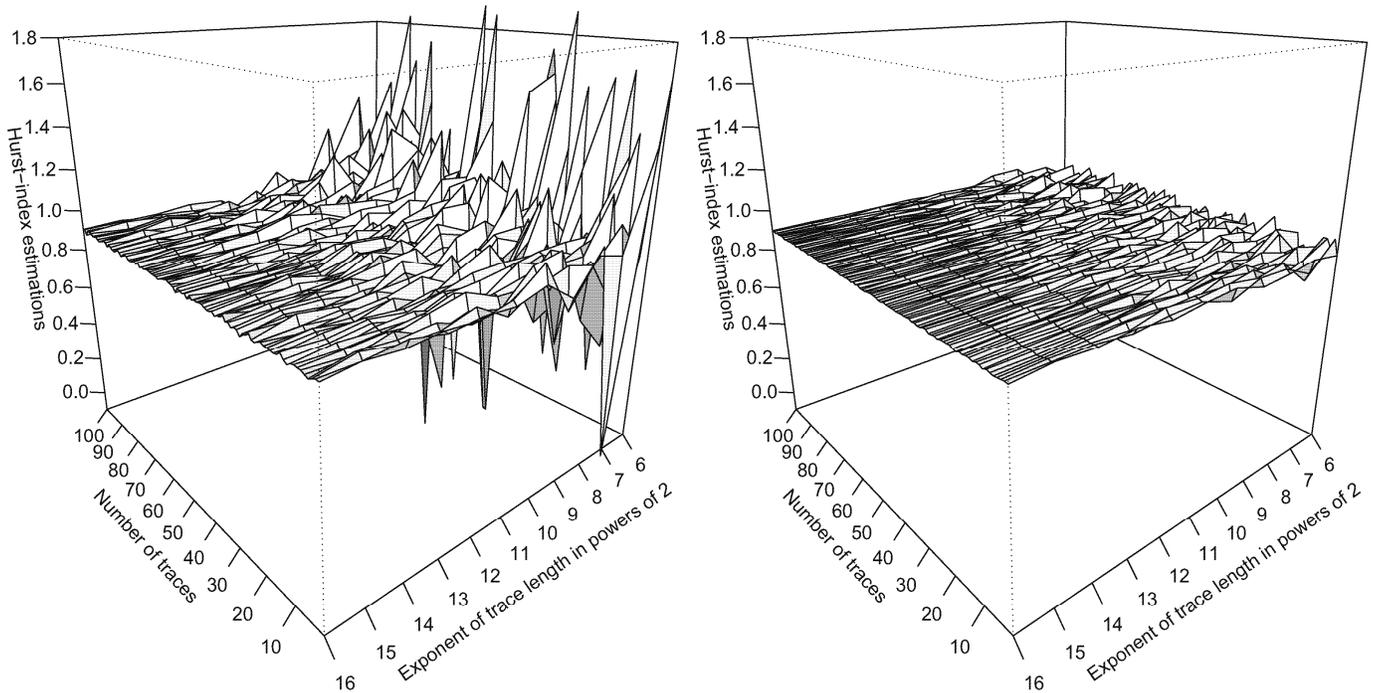

Figura 1. Estimaciones de wavelet y de Whittle para 100 trazas fGn con $H = 0.81$. El gráfico de la izquierda corresponde a las estimaciones utilizando wavelet y el de la izquierda a las estimaciones de Whittle.

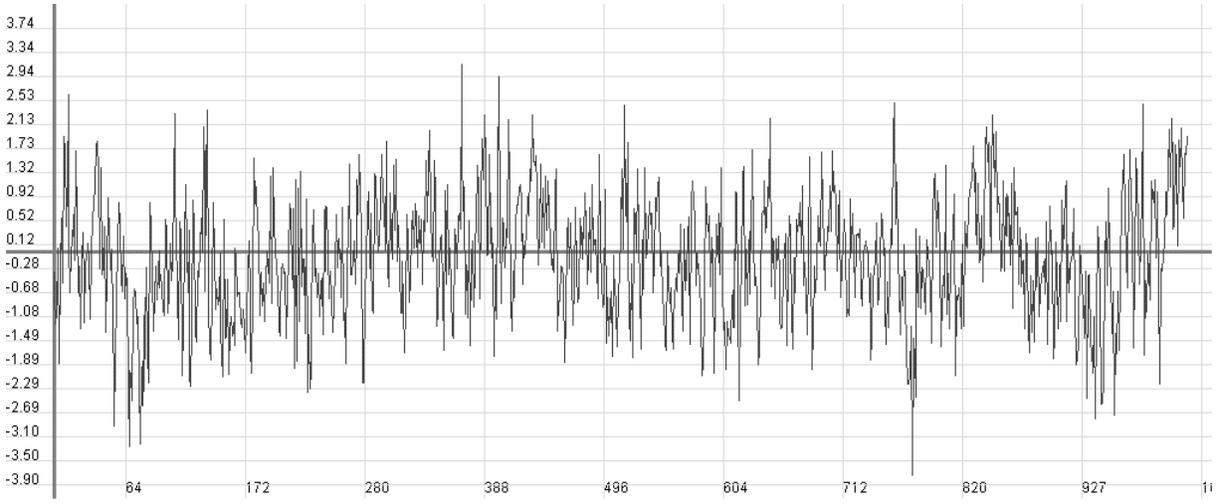

Figura 2. Ejemplo de una de las 100 trazas sintéticas fGn con $H = 0.81$ empleadas en la Fig. 1 desarrollada con SELFIS [15].

Cuando $2^{11} < N < 2^{12}$ las estimaciones muestran resultados aceptables y cuando $N \geq 2^{13}$, las estimaciones son altamente precisas. Se obtiene un comportamiento similar para trazas con valor de Hurst-index diferente que $H = 0.90$. El mismo tipo de gráfica cuando se usan estimaciones tipo Whittle se muestra en la parte derecha de la figura para las trazas donde $H = 0.90$. Como se puede ver, en general, las estimaciones de Whittle presentan alta precisión y baja variabilidad para series temporales cortas con LRD. Para los valores de $H$ por debajo de 0.90, el índice de Hurst estimado presenta alta variabilidad cuando $N < 2^{10}$ y $H \geq 0.90$ Generalmente, las estimaciones son precisas considerando $N > 2^9$.

Comparando los resultados según el estimador de Whittle con los de los basados wavelets, resulta Whittle más preciso a razón de series fGn de corta longitud.

La Figura 3 expone lo anterior desde la perspectiva de las estimaciones del parámetro de Hurst para trazas con $H = 0.90$ y longitudes variables al aplicar el método del periodograma y el estadístico R/S. En este sentido se recuerda que el método del periodograma basa su comportamiento en las cercanías de la PSD de la serie bajo análisis. Luego, como se observa en la parte izquierda de la figura, el comportamiento analizado con el método del periodograma se asimila al método wavelet para series cortas, es decir, con sesgo y variabilidad significativas. Un comportamiento similar se observa al utilizar valores de $H$ diferentes. El gráfico correcto de la figura también muestra el mismo tipo comportamiento descrito según el periodograma y el estadístico R/S. Téngase en cuenta que en la gráfica que la R/S presenta una alta variabilidad sin importar la longitud de las trazas. La variabilidad disminuye cuando las longitudes de las trazas aumenta, pero se complejiza a la vez establecer una longitud mínima para estimaciones precisas de *H*. En contraste con los otros métodos expuestos, la superficie en el diagrama de perspectiva R/S es siempre áspera. Luego, los resultados de las simulaciones dan cuenta de que el estadístico R/S tiende a un comportamiento sesgado con independencia de la cantidad de puntos. Por otra parte, la Fig. 4 expone el comportamiento del sesgo para todos los métodos estudiados para *H* = 0.90.

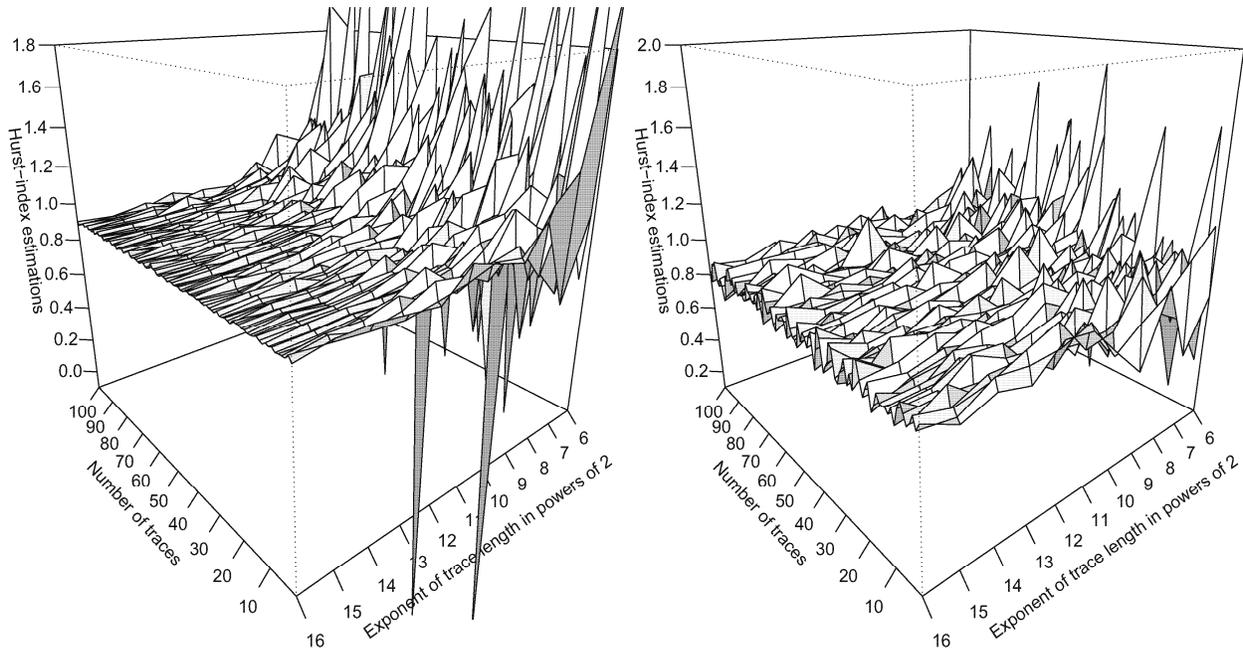

Figura 3. Estimaciones de periodograma y estadístico R/S para 100 trazas del tipo fGn con *H* = 0.81. El gráfico de la izquierda corresponde a las estimaciones utilizando periodograma y el de la izquierda a las estimaciones utilizando el estadístico R/S.

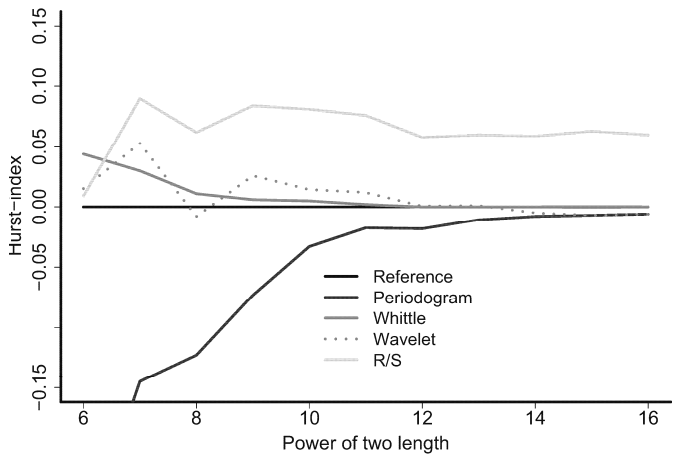

Figura 4. Sesgo para todos los métodos.

De las Fig. 1 y 3 se observa que los métodos de Whittle y wavelet son los con mejor comportamiento para series cortas. A diferencia de wavelet, el método Whittle se comporta con menos irregularidad para series cortas y para $N \geq 2^8$ los sesgos no son significativos. Para los métodos de wavelet, el sesgo es irregular para las series cortas y se estabiliza en $N \geq 2^{12}$. Los otros métodos (ver en Fig. 3 estadístico R/S y periodograma) exhiben un comportamiento por completo irregular y un sesgo muy alto; hasta $N \leq 2^{16}$ sus estimaciones no son aceptables. El sesgo del método con el estadístico R/S parece no tener un comportamiento estabilizador, mientras que el periodograma parece estabilizarse para *N* elevado ($\geq 2^{16}$). La Figura 5 ilustra el comportamiento de las desviaciones estándar para las trazas con *H* = 0.90 y longitud variable *N*. El método Whittle es el estimador que presenta menos variabilidad y para $N \geq 2^8$, las estimaciones son lo suficientemente precisas. Los métodos de wavelet y periodograma son los que siguen en precisión y la longitud requerida es relativamente la misma cuando se toma en cuenta la variabilidad. Sin embargo, considerando el sesgo y la desviación estándar juntos indican que el mejor estimador para series temporales cortas es el método Whittle. El método de Whittle presenta alta precisión cuando $N \geq 2^8$, el método wavelet presenta una buena precisión si $N \geq 2^{13}$, estimaciones aceptables del periodograma cuando $N \geq 2^{15}$ y el método R/S exhibe estimaciones sesgadas con $N \in [2^6, 2^{16}]$.

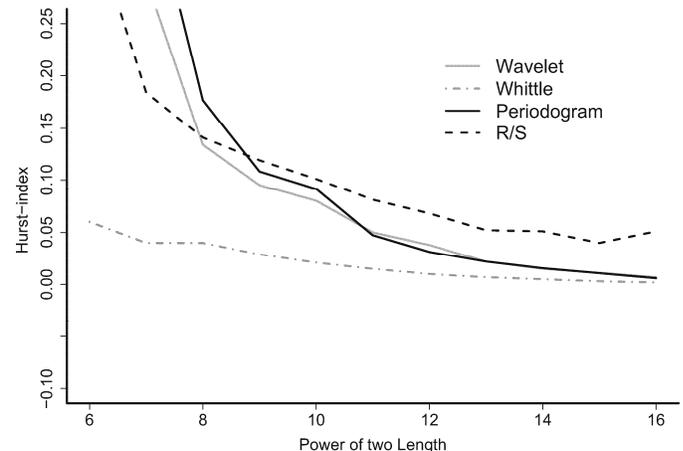

Figura 5. Desviación estándar para todos los métodos.

Para cada serie de longitud $M = 2^{16}$ se realiza una variación del análisis convergente definido anteriormente. El análisis se aplica primero a los primeros $\tau_0 = 2^6$ puntos de las series $X_j$, es decir, se aplica el método de estimación de $H$, $\hat{H}_{t_i}^{N_{min}}(\cdot)$, a los primeros $\tau_0$ puntos de $X_j$. Luego se aplica repetidamente $\hat{H}_{t_i}^{N_{min}}(\cdot)$ a los siguientes $\tau_0 + i\tau_u$ puntos de $X_j$, donde $\tau_u = 200$ e $i = 1, 2,\ldots, k$ tal que $\tau_0 + k\tau_u \leq M$. Este análisis se realiza a 100 series fGn obteniendo así el comportamiento convergente de cada una. Una vez realizado el análisis de convergencia para cada una de las trazas estudiadas, se realiza el análisis de convergencia promedio, esto significa la media para las 100 estimaciones de $\tau_0$ y así sucesivamente. La gráfica de convergencia media $\hat{H}_{t_i}^{N_{min}}(\cdot)$ versus $\tau_i$, $i = 0, 1,\ldots, k$ es ahora un indicador de qué tan bien convergen los estimadores al valor teórico de $H$. Este comportamiento convergente medio se aplicó a los estimadores estudiados en este artículo. La Fig. 6 muestra el comportamiento convergente medio del método estadístico R/S. Observe que el estadístico R/S se estabiliza rápidamente pero está sesgado por 0.05, por lo que es difícil proponer una longitud mínima. La Fig. 7, Fig. 8 y Fig. 9 muestran el mismo tipo de gráficas de convergencia para los otros métodos estudiados. Tenga en cuenta que el método de tipo Whittle se estabiliza rápidamente con un sesgo muy bajo. El comportamiento es similar al del estadístico R/S, pero a diferencia de la estadística R/S, las estimaciones en Whittle son muy poco sesgadas. El comportamiento del periodograma y el método wavelet son similares. El periodograma, según la definición dada en la Sección III, está sesgado negativamente, mientras que wavelet está en principio positivamente sesgado, pero a largo plazo se vuelve sesgado negativamente. De la Fig. 7 se infiere que para una estimación precisa del índice de Hurst, se necesitan al menos $2^9$ puntos para el método de tipo Whittle. Para el método wavelet basado en la Fig. 8, se infiere que se necesitan al menos $2^{12}$ puntos para estimaciones precisas y para el método del periodograma (Fig. 9) se necesitan a lo menos $2^{13}$ puntos para lograr estimaciones precisas. De la Fig. 6 es necesario rescatar que para análisis visuales el método R/S sigue siendo apropiado.

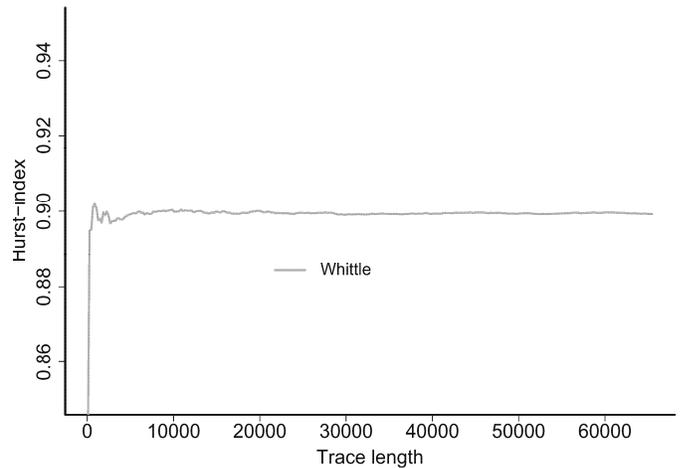
Figura 7. Convergencia media del método Whittle.

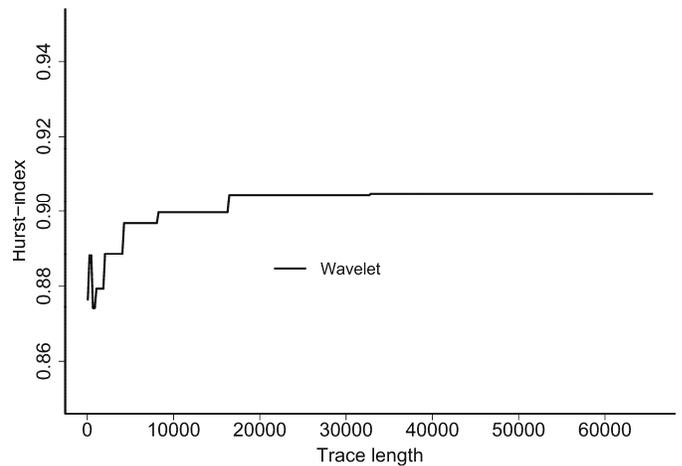
Figura 8. Convergencia media del método wavelet.

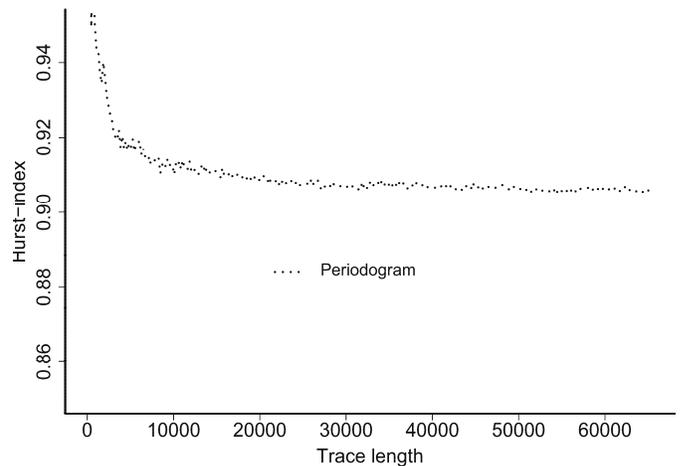
Figura 9. Convergencia media del método del periodograma.

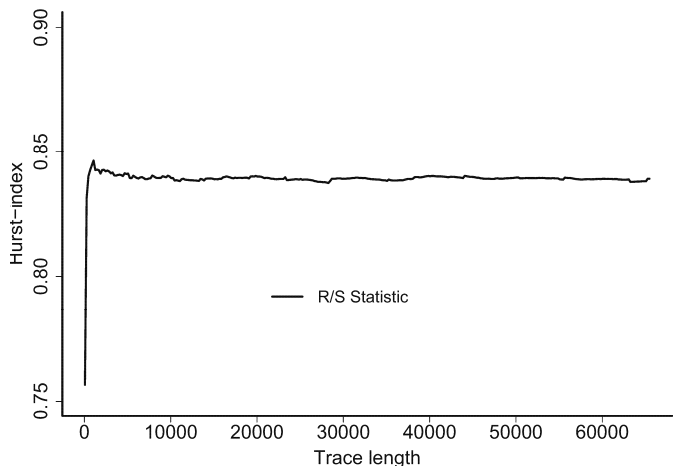
Figura 6. Convergencia media del estadístico R/S.

## V. Aplicación a Trazas de Tráfico Reales

Ahora se aplican los resultados anteriores a trazas reales de tráfico. La traza seleccionada es la traza Ethernet de Bellcore medida en agosto de 1989 [29]. La traza representa una hora d tráfico LAN. El análisis realizado es el mismo explicado en la sección III, sea $I_x = t_0, t_1,\ldots, t_k$ una secuencia de puntos en el eje $x$, donde $t_{i+1} > t_i$ y $(t_{i+1} - t_i) = 2^8$, a cada bloque de la traza

$X_j$ de longitud $2^8$, $\{X_j\}_{j=t_i}^{t_i+256}$, se aplica el método de estimación de $\hat{H}_{t_i}^{N_{min}}(\cdot)$ y se construye el gráfica del comportamiento del estimador $\hat{H}_{t_i}^{N_{min}}(\cdot)$. La Fig. 8 muestra los resultados de este análisis. Se observa que el periodograma sobrestima y el estadístico R/S muestra un comportamiento irregular. El estimador Whittle sigue el valor $H$ reportado en [17], [18] y el método basado en wavelet sigue el valor reportado con alta variabilidad. Entre las posibles áreas de aplicación de los resultados se encuentran las series de tiempo fisiológicas en las que se obtienen series temporales de corta duración [10], [11], la administración de parámetros de QoS en tiempo real, donde se requiere una breve medición para tomar decisiones de desempeño y en cada disciplina en la que la duración de la serie temporal sea un factor que afecta el rendimiento de un sistema [30].

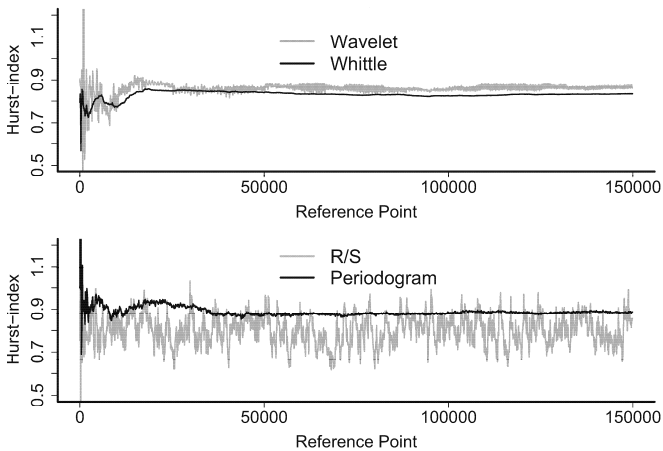

Figura 8. Comportamiento de la traza LAN en el tiempo.

## VI. Conclusiones

Los modelos de tráfico tradicionales basados en procesos con dependencia de corto alcance, no describen con detalle el comportamiento de los flujos en las actuales redes de datos de alta velocidad. Consecuentemente, es necesario replantear el estudio de los modelos de los sistemas de carga de las redes de computadoras considerando tráficos de entrada autosimilares, puesto que sus requerimientos imponen nuevos desafíos a la ingeniería de redes en especial en estrategias de buffering para equipos activos y estimación de rendimientos.

Este trabajo presentó un estudio del comportamiento de los estimadores en series cortas en el contexto de trazas de fGn.

Sobre la base del estudio del comportamiento de miles de series temporales de fGn bajo sesgo, $\sigma$, análisis convergente y comportamiento para una longitud determinada $N_{min}$, supuesta como longitud mínima se llega a las siguientes conclusiones.

El método Whittle se comporta de mejor manera para series de cortas y largas que presentan tanto sesgo como variabilidad mínimos. Los métodos Wavelet y de periodograma exhiben un comportamiento aceptable con series de longitud media. De la misma manera el estadístico R/S se comporta con un sesgo alto y no es adecuado para mediciones de corta longitud.

La longitud mínima para estimar $H$ se propuso para todos los métodos anteriores. Sobre esta base la longitud mínima del estimador Whittle es al menos $2^8$, $2^{13}$ para el método wavelet y $2^{15}$ para un periodograma. Así mismo no es posible un reporte para el estadístico R/S debido a su alto sesgo y variabilidad en longitudes mayores a $2^{16}$ puntos.